# Study of Water Speed Sensitivity in a Multifunctional Thick-film Sensor by Analytical Thermal Simulations and Experiments

F. Stefani[1], P.E. Bagnoli[1], S. Luschi[2]

(1) Dept. of Information Engineering, University of Pisa, Via Caruso 16, I-56100 Pisa, Italy, Email : fabio.stefani@iet.unipi.it, p.bagnoli@iet.unipi.it

(2) AMIC srl, Via delle Cateratte 94/C, 57122 Livorno, Italy, Ph +39 0586 340053, Fax +39 0586 340056, Email : stefano.luschi@amicweb.com

*Abstract-* A multifunctional (temperature, liquid flow, pressure and electrical conductivity) thick film sensor for monitoring water pipelines is here presented. This work is mainly focused on the theoretical and experimental characterization of the water flow sensitivity based on a planar version of the well-known hot-wire anemometer. The simulations of the temperature displacement on the sensor surface under several electrical biases and heat convection conditions were performed by means of the fast analytical thermal simulator DJOSER, thus providing an example of the capability and the utility of this simulation program. The calculated sensitivity curves to the heat convection coefficient and/or to the water speed were found to be in agreement with the experimental data measured on the sensor mounted within a closed pipeline which allows changing the water speed until 1000 liters/hour.

## I. INTRODUCTION

Drinkable water is a more and more precious resource for the human society. Its importance requires the monitoring of the public water pipelines in order to prevent, as much as possible, leakages and wasting and to *in situ* measure the physical parameters for controlling the liquid quality. For this reason a smart multifunctional sensor for the public water pipelines was recently designed and patented by AMIC s.r.l. of Livorno (Tuscany) and University of Pisa.

Beside it was realized in not up-to-date and not miniaturized technology, i.e. thick-film hybrid technology on alumina substrate, it offers the advantage to provide a complete set of measurements for water monitoring in a compact form. In fact it integrates four different sensitivities to water temperature, pressure, electrical conductivity and liquid flux (or water speed up to 10000 litres/hour) within the pipeline and it contains aboard part of the electronic circuits for the sensing signals conditioning. The water speed sensitivity, on which the present paper is mainly focused, is based on a planar version of the<hot-wire> anemometry [1,2,3]. Two different thick film resistors were printed on the sensor top surface: Rh (hot

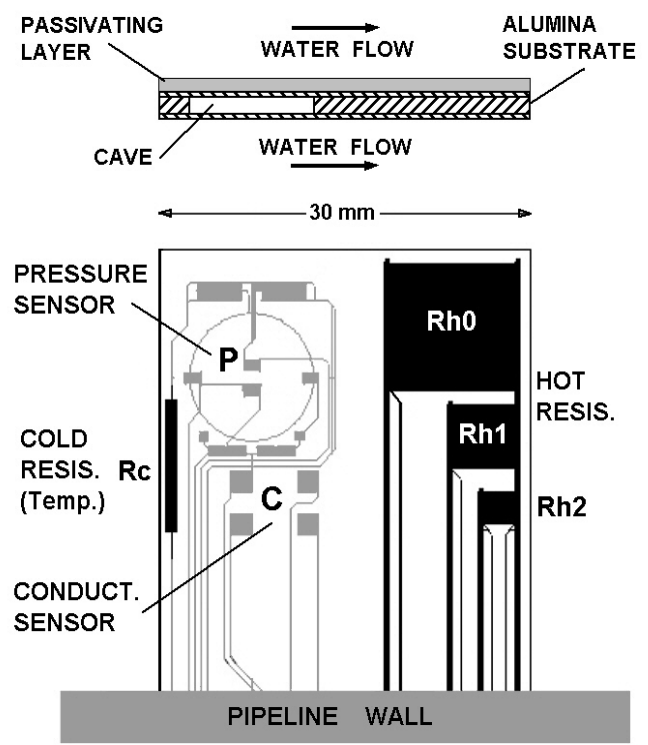

Fig. 1. Top view and cross-section of the multifunctional sensor. Rc is the temperature sensor; Rh0-Rh2 are the hot resistors for liquid flow measurements; the groups P and C are the sensors for pressure and water conductivity respectively.

resistor)with a large area and driven at high electrical power and Rc (cold resistor) with a negligible power dissipation. Both these two devices, exchanging heat with the surrounding water by convection, are temperature sensitive by means of a known law so that the respective temperatures (Th and Tc respectively) can be electrically measured.

Due to the heat exchange by convection, the value of Th depends on the heat convection coefficient (h) which is an increasing function of the water speed. However the forming feedback circuit connected to the sensor, using the measured temperature values in the two resistors, is able to keep the temperature difference constant by changing the electrical power (P) in the hot resistor. This power may be used to indirectly evaluate the water speed (or the water flux within the pipeline) since it is an increasing function of the heat convection coefficient (h).

Since the sensitivity to water speed is directly related to thermal effects, the analytical thermal simulator DJOSER [4, 5, 6] for multilayer structures in the steady state regime was used to predict the sensitivity behaviour of the system as a function of the convection heat exchange coefficient, under a wide range of thermal boundary conditions and electrical bias and to investigate the contribution of structural details as the thermal properties of the substrate and the passivating layer [7]. The sensitivity properties of the sensor were also experimentally recorded as a function of the water flux or liquid speed and the experimental data were compared with the calculated ones by extrapolating a relationship connecting the convection coefficient with the liquid speed within the pipeline in the turbulent dynamic regime.

## II. SENSOR STRUCTURE

The figure 1 shows the top view and the cross- section (top of figure) of the multi functional sensor. The structure is composed by a thick alumina substrate (800 microns) having a circular hole and covered on the top and bottom faces by a thinner (100 microns) alumina foils. This multilayer structure is designed in order to create a cave connected to the external atmosphere providing a reference for the pressure sensor.

The thick film resistors and the gold metal interconnections were printed on the top of the structure and covered by a thermally conductive plastic (50 microns thick) as a passivating layer separating the electrical resistances from the surrounding water.

The narrow resistance Rc, also addressed as 'cold resistor' and whose value is 200 ohm at 25 °C, is the water temperature sensor. In fact, it is driven with a current not exceeding 80 microamperes so that the power dissipation within this resistor may be assumed as negligible and the resistor may be retained in thermal equilibrium with the surrounding water. The resistance Rc is measured using the two-wire technique.

The four resistances printed on the inner circular cave (group P in Fig. 1) are connected in a resistive bridge which is balanced if the cave internal and external pressures are equal. If the water pressure increases with respect to the external atmosphere, the mechanical deformation of the upper membrane causes the unbalancing of the bridge providing an electrical signal which has a linear behaviour with the water pressure.

The group C in Fig. 1 is composed by four metal (platinum) contacts used for the water electrical conductivity measurements using an alternate voltage signal. The platinum pads are not covered by the upper passivating layer and are in direct contact with the liquid. The resistors drawn in black colour in Fig. 1 are used to measure the water speed.

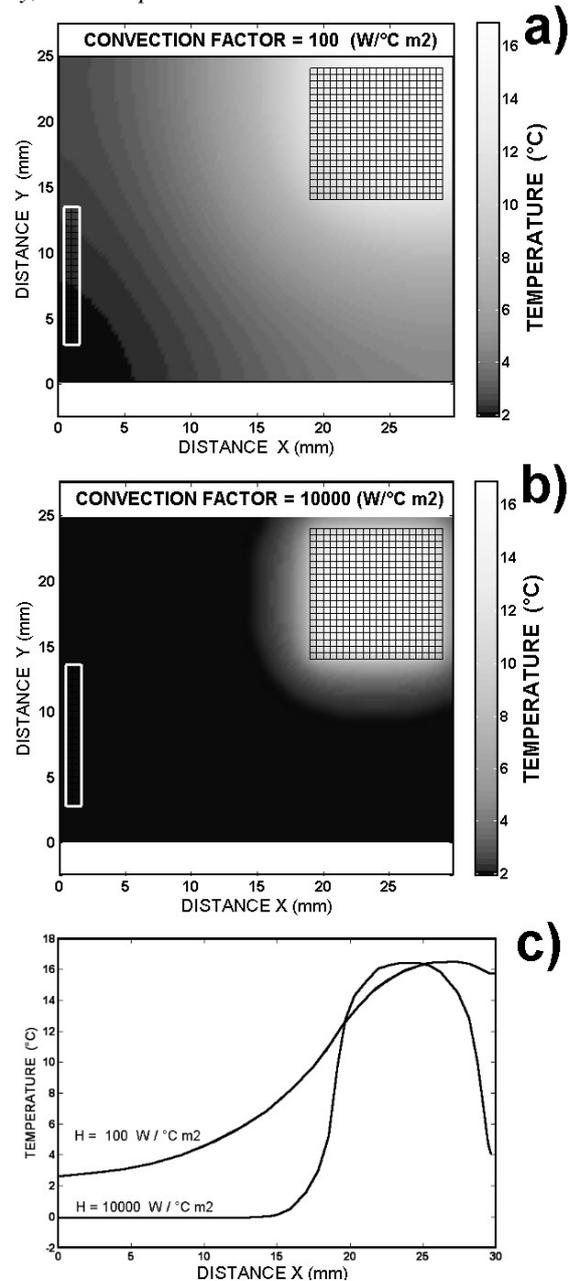

Fig. 2. Temperature maps calculated for the power dissipated on the Rh0 resistor and with a temperature difference of 16 °C: a) the convection coefficient of 100 W/°C m². b) ) the convection coefficient of 10000 W/°C m². c) Temperature plots along the horizontal axis crossing the centre of the Rh0 resistor.

The three hot resistors (Rh0, Rh1 and Rh2, all having a resistance value of 5 ohm) with square shape and different areas were printed close to the right border of the substrate in order to increase as much as possible their distance from the cold resistor Rc. This was done for preventing the cross-heating induced on the cold resistor by the hot ones, also prevented by the relative position of Rc and Rh with respect to the direction of water flow. As can be seen from the metal contacts to the hot resistors, their resistance values are measured using the four contacts methods. In the showed prototype of the multifunctional sensor, the different areas of the three hot resistors were used to test the influence of this parameter on the water speed sensitivity.

## II. THERMAL SIMULATIONS

The thermal flux P exchanged by convection by a planar and uniform surface in contact with a liquid is generally ruled by the well-known relationship:

$$P = h \, A \, (Th\text{-}Ta) \qquad (1)$$

where A is the dissipating area and Ta is the liquid temperature. From this equation it follows that, if the temperature difference (Th-Ta) is kept constant, the power P is linearly proportional to the convection heat exchange coefficient h. However, due to the geometrical factors, to the presence of a passivating layer interposed between the power dissipating areas and to he heat flux path through the sensor body, the P(h) relationship is far to be linear and a proper thermal simulation tool must be used in order to predict this sensitivity function.

The thermal simulator DJOSER, thanks to its speed and modular software structure, was used for calculating the temperature difference (Th-Tc) in the steady-state condition under a given power dissipation within the hot resistor and as a function of the h parameter.

Furthermore, part of the DJOSER program, i.e. the algebraic system solver, was used to implement a feedback calculation strategy allowing calculating the power dissipation able to keep the temperature difference between the hot and cold resistors at a given value. This strategy practically reproduces the feedback control operated by the electronic circuits for sensor signal conditioning.

The DJOSER model of the sensor includes the three alumina layers of the substrates, the thermal contact resistances of the gluing interfaces and the top passivating layer. As a first approximation the presence of the circular inner cave of the pressure sensor was neglected. Furthermore since the width of the structure is much smaller than the pipeline diameter, the heat exchange coefficient with the liquid was retained as uniform on the upper, lower and lateral surfaces except the lateral face connected with the pipeline wall which is set as adiabatic. For each thermal simulation step the temperatures Th and Tc were calculated as the average temperature values over the areas of Rh and Rc respectively.

Figure 2 shows in grey scale two temperature maps of the sensor top surface under two different values of the convection coefficient and for the same average temperature difference (16 °C) between the hot (Rh0) and the cold resistors whose areas are marked in the two maps by the rectangles. Fig 2a is referred to a convection coefficient of 100 W/°C m$^2$ while the h value for the map of Fig. 2b was 10000 W/°C m$^2$.

The temperature plots along the horizontal axis crossing the centre of Rh0 are compared in Fig. 2c. The effects of the heat exchange by convection with the surrounding fluid on the sensor thermal conditions under operating regime are evident from the above graphs: the temperature displacement is strongly dependent on the convection conditions. For higher value of the coefficient, the temperature rise above the ambient one is practically concentrated in the area of the hot resistor, while for lower values, and hence for lower liquid speeds, the heating of the sensor surface tent to be more diffused.

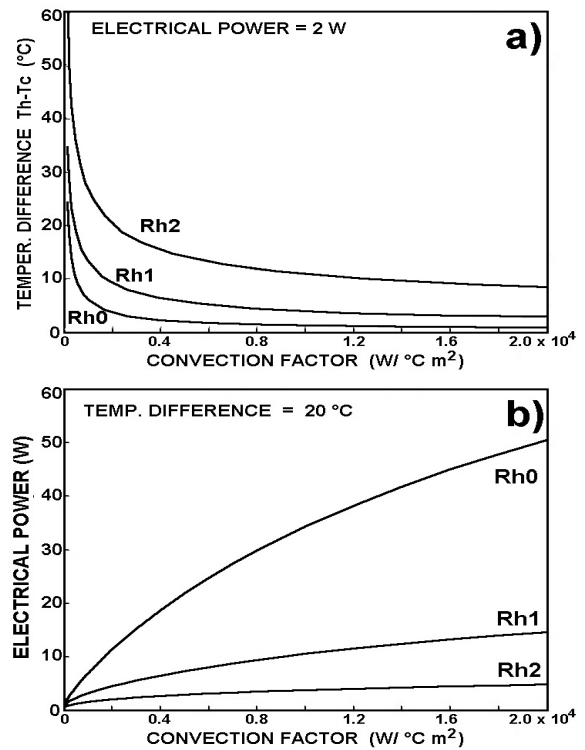

Fig. 3. Calculated sensitivity curves for the three hot resistors as a function of the convection coefficients. a) The plots are referred to constant power dissipation. b) The plots are referred to a constant temperature difference between the hot and cold resistors.

The sensitivity curves as a function of the convection coefficient were calculated for the three hot resistors and for both the operating conditions [8, 9, 10] which are at constant electrical dissipated power and at constant temperature difference between the two resistors. The obtained plots are shown in Fig 3a and Fig. 3b respectively.

By observing the plots in Fig. 3b, it should be noted the strong influence of the resistor area so that this parameter can be used just for setting the sensor sensitivity. However, it is quite obvious that the slope of the curves is an increasing function of the temperature difference too. Furthermore the different thermal displacement as a function of convection coefficient, as above shown, seems to be the main cause of the lack of linearity of the sensitivity curves at constant temperature difference and of their disagreement with respect of what the relationship (1) foresees.

The analytical thermal simulations performed with the program DJOSER were also used for investigating the dependence of the sensitivity curves on the physical and structural parameters of the designed sensors. In particular, the thermal characteristics of the upper passivating layer were found to play a crucial role. In fact, just at the top surface where the convection heat exchange must be as large as possible, this layer inserts a thermal resistance between the resistor and the liquid, therefore causing a worsening of the heat exchange by convection.

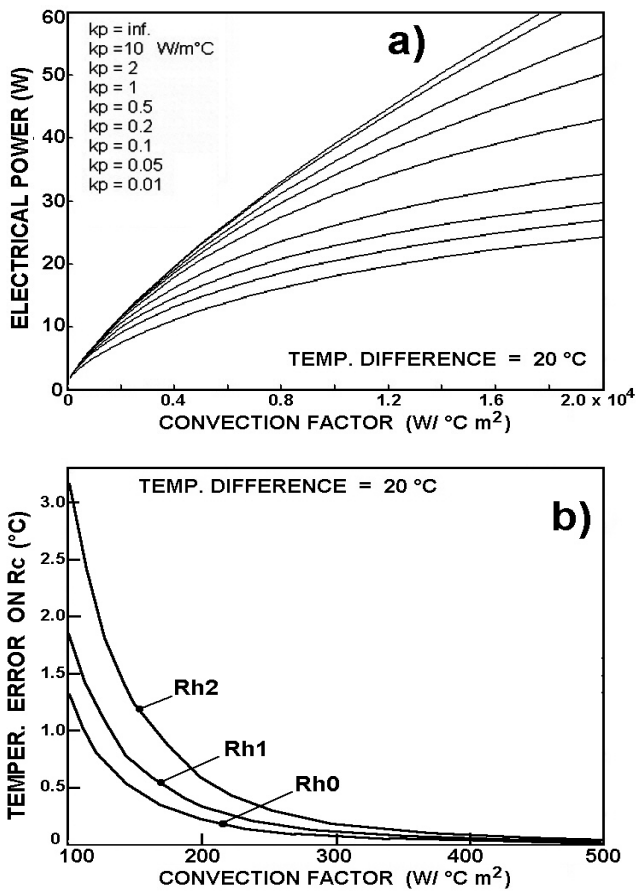

Fig. 4. a) Sensitivity curves calculated for Rh0 and a temperature difference of 20 °C for several values of the thermal conductivity of the top passivating layer. b) Curves of the error in the temperature reading of the cold resistor induced by the hot resistor as a function of the convection coefficient.

The figure 4 a shows several plots for the Rh0 resistor, each calculated using a different value of the passivating layer thermal conductivity whose thickness was supposed to be 50 microns. From these plots it is quite evident that, as the thermal conductivity of the passivating layer decreases, the sensitivity curve undergoes both attenuation and a loss of linearity. The actual value for the insulating layer thermal conductivity, which is 0.2 W/ m °C being epoxy plastic, is far to be the optimal choose. A charged glass material with higher conductivity (about 3 W/ m °C) should be preferred if the problems connected with the insulating capability against water and thermo-mechanical stability will be solved.

The figure 4b shows the thermal cross-talk between the cold and the hot resistors. It means that the power dissipation within one of the hot resistors can cause a heating of the first one just due to the heat conduction across the alumina substrate. Since the cold resistor is also used as a liquid temperature sensor, this thermal cross-talk induces an error in the temperature reading.

The plots of Fig. 4b, calculated as a function of the convection coefficient and for the three hot resistors respectively, clearly indicate that this effect is quite negligible except for the lower range of the parameter h, close to the condition of quiet liquid (zero water flux). Furthermore, it is also shown that, at least for the operating condition chosen i.e. constant temperature difference, the thermal cross-talk is higher for the resistors with lower area. This last statement together with the increased sensitivity to the h parameter indicates that a larger area of the hot resistor must be preferred for the best sensing performances.

## II.  EXPERIMENT

The experimental tests on the water speed sensor were carried out mounting it on the wall of a closed loop pipeline as shown in Fig. 5. A water pump inserted in the loop allows changing the liquid flux up to 1200 litres/hour. The sensor under test was inserted at a distance of about 5 diameters from the outlet of a plenum chamber in order to provoke a stabilized turbulent flux in the cross-section where the sensor was placed. Both the water flux and water temperature were separately measured by means of a standard flux meter placed just before the chamber and by an insulated thermocouple placed after the sensor. The inner side of the pipeline was 42.5 millimetres.

Figure 6 schematically shows the computer controlled measurement apparatus connected to the external pins of the multifunctional sensor. The electrical power to the hot resistor was given with an alternate voltage waveform generated by a waveform generator at a frequency of 1000 Hz. The choose of using sinusoidal signals instead of continuous voltages was induced by the needs of preventing any electrical polarization of the resistors which, in presence of small leakages in the insulator layer, might cause the birth of electrolytic small bubbles on the sensor surface and therefore a worsening of the convection heat exchange conditions. The power of the voltage signal is granted by a power amplifier. A multimeter is connected to the sensor outputs in order to measure the resistance values of the cold and the hot resistors and the thermocouple signal for the water temperature measures.

The measuring procedure was ruled by the personal computer which uses the LabView software. Two different operating procedures may be performed: a) constant power dissipation and b) constant temperature difference. In this last operating regime the electrical power to the hot resistor is continuously corrected by the software program in order to keep the temperature drop at a given value previously set. Due to the speed of the computer controlled feedback the system is able to reach the steady-state conditions in a minute after an abrupt change of the water flux.

The typical experimental results for the constant temperature difference regime are shown in Fig. 7. Here the electrical power versus water flux plots are traced for the hot resistor Rh1, for three values of the temperature difference and in the range up to 1000 litres/hour. For sake of clarity, the abscissa axis was labelled also with liquid velocity scale (meters/ seconds). As can be seen the general trend of the experimental plots is fully in agreement with the behaviour predicted by the thermal simulations which is a non linear shape with a downward concavity. Furthermore, as was expected, the electrical power on the hot resistor is perfectly proportional to the temperature difference for each water flux value.

### III. CALIBRATION PROCEDURE

The input parameter needed by the thermal simulator is the convection coefficient h while that of the experiments is the water speed (v) or water flux. In order to compare the theoretical data with the experimental ones a proper relationship between the average water speed within the pipeline and the convection coefficient must be established [11]. This calibration procedure, which was performed for the data measured under constant temperature difference and in the range 0 - 1000 liters/hour, is based on the assumption that a given power value in the hot resistor was obtained in the simulation system using given h value, while in the experimental characteristics it corresponds to a liquid speed v. Therefore, since the power value is the same, the h coefficient must be generated by the speed v.

At first, the P(v) function was obtained by fitting the data from the experimental measurements using the empiric law for the hot-wire anemometry, addressed as the King's law [1, 12, 13], which is the following

$$P(v) = dT\,(a + b\,v^n), \qquad (2)$$

where dT is the temperature drop and a and n, whose typical value is 0.5, are suitable parameters. The best fitting of the curves yielded the following values a=0.13525; b=0.3325; n=0.593; with $R^2$=0.9916.

The P(h) curve, as obtained from the simulation results, may be fitted using a polynomial function which comes from a simple resolution of an equivalent thermal network containing all the thermal resistances and convection coefficients on all the side of the structure as concentrated parameters. The fitting function used has the following form:

$$P(h) = dT \cdot \frac{a \times h^2 + b \times h}{c \times h^2 + d \times h + 1} \qquad (3)$$

The best fitting gave the following values for the four parameters a=2943, b=896e+03, c=1188, d=1.11e+07; with of $R^2$=0.9994.

The P(h) and P(v) functions, being monotonic, can be inverted in order to obtain the reverse functions h(P) and v(P) for the given temperature difference and for a power range from 0 to 5 W. The obtained values of the function, organized in a look-up table, can be used to convert the convection coefficient in water speed and *viceversa*. An analytical direct relationship h(v) may be also obtained by equating the P(h) and P(v) function. Fig 8a shows the plot of the convection coefficient as a function of the average water speed obtained using the above described procedure.

In order to test the reliability of this calibration curve, it was compared with a fully theoretical function which connect the Nu (Nusselt), Pr (Prandt) and Re (Reynolds) numbers [14] for water in a turbulent dynamic regime and for the present geometry of the pipeline. This function may be written in a compact form as follows.

$$h = a + b*(v^n) \qquad (4)$$

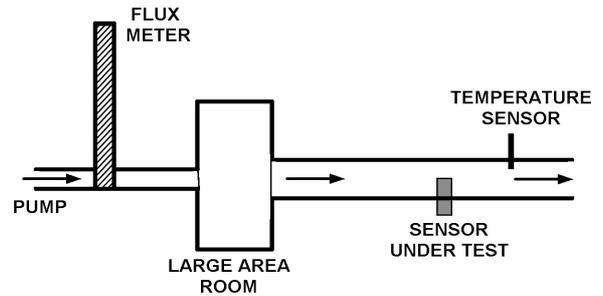

FLUX METER

PUMP

LARGE AREA ROOM

TEMPERATURE SENSOR

SENSOR UNDER TEST

Fig. 5. Experimental setup for the sensor characterization

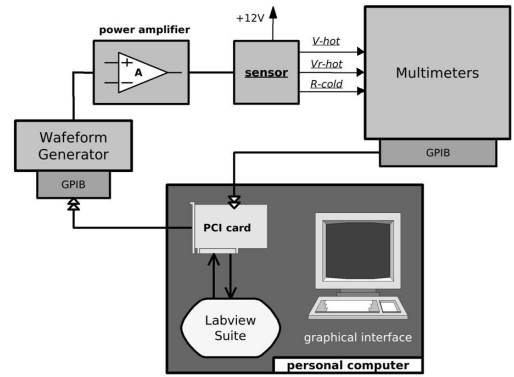

Fig. 6. Measurement setup for the sensor characterization

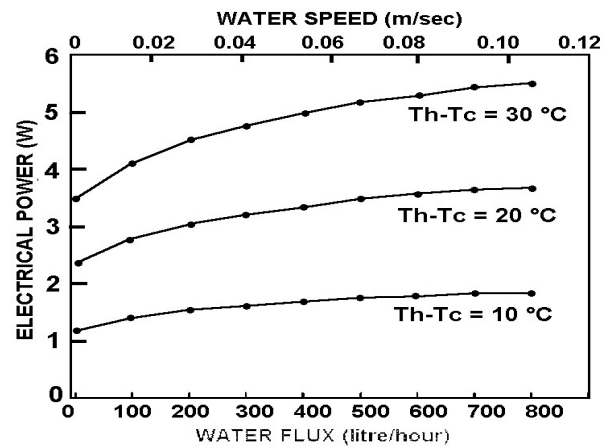

Fig. 7. Experimental plots for Rh1 hot resistor as a function of the water flux and liquid speed and for three different values of the temperature difference between the hot and cold resistors.

where the exponent n is 0.5 for a laminar flow and 0.8 for a turbulent one. In the present case we used the following values for the three parameters: a=242.3, b=1443, n=0.7191

The agreement between this last function and the previous one is shown in the Fig. 8b. Here the relative error percentage between the two functions is plotted versus the water speed. As can be seen the error is close to zero except in the beginning speed range in which its value does not exceed 2 %. This last behavior is probably caused by the change of the flux dynamics that becomes laminar for very low speed values notwithstanding the proximity of the plenum chamber.

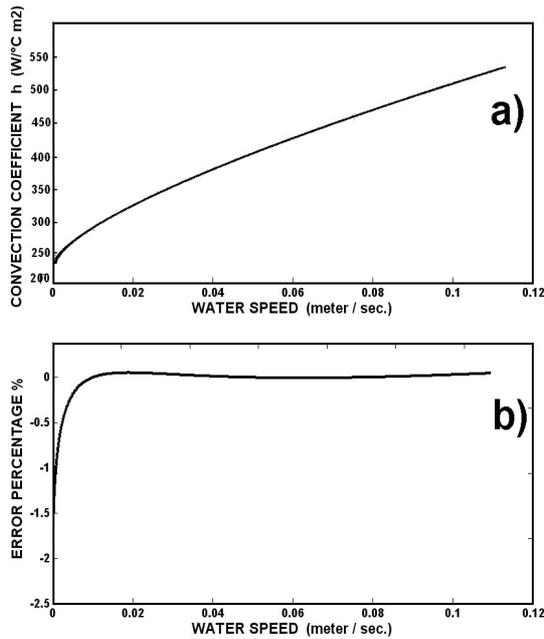

Fig. 8. a) Estimated relationship between the convection coefficient and the water speed. b) Relative percentage difference plot between the above curve and a theoretical relationship describing the same function in the turbulent regime.